\documentclass{emulateapj}
\pdfoutput=1

\usepackage{hyperref}
\hypersetup{
pdftitle={Low mass nearby Hills stars in Stripe 82},
pdfauthor={Yanqiong Zhang},
colorlinks=true,
linkcolor=blue,
citecolor=blue,
urlcolor=blue
}

\newcommand\kms{$\rm km~s^{-1}$}

\shortauthors{Zhang, Smith \& Carlin}
\begin{document}

\title{Red Runaways II: Low mass Hills stars in SDSS Stripe 82}
\shorttitle{Red Runaways II}

\author{Yanqiong Zhang$^{1,2}$, Martin C. Smith$^1$, Jeffrey L. Carlin$^{3,4}$}
\affil{$^1$Key Laboratory of Galaxies and Cosmology, Shanghai Astronomical Observatory, Chinese Academy of Sciences, 80 Nandan Road, Shanghai 200030, China}
\affil{$^2$Graduate University of the Chinese Academy of Sciences, No.19A Yuquan Road, Beijing 100049, China}
\affil{$^3$Rensselaer Polytechnic Institute, Troy, USA}
\affil{$^4$LSST and Steward Observatory, 933 North Cherry Avenue, Tucson, AZ 85721, USA}
\email{zhangyq@shao.ac.cn}
\email{msmith@shao.ac.cn}

\begin{abstract}
Stars ejected from the Galactic centre can be used to place important
constraints on the Milky Way potential. Since existing hypervelocity
stars are too distant to accurately determine orbits, we have
conducted a search for nearby candidates using full three-dimensional
velocities. Since the efficacy of such studies are often hampered by
deficiencies in proper motion catalogs, we have chosen to utilize
the reliable, high-precision SDSS Stripe 82 proper motion catalog.
Although we do not find any candidates which have velocities in excess of
the escape speed, we identify 226 stars on orbits that are consistent
with Galactic centre ejection. This number is significantly larger
than what we would expect for halo stars on radial orbits and cannot
be explained by disk or bulge contamination. If we restrict ourselves
to metal-rich stars, we find 29 candidates with $\rm [Fe/H] > -0.8$
dex and 10 with $\rm [Fe/H] > -0.6$
dex. Their metallicities are more consistent with what we expect for
bulge ejecta, and so we believe these candidates are especially
deserving of further study.
We have supplemented this sample using our own radial velocities,
developing an algorithm to use proper motions for optimizing candidate
selection. This technique provides considerable improvement on the
blind spectroscopic sample of SDSS, being able to identify candidates
with an efficiency around 20 times better than a blind search.
\end{abstract}
\keywords{Galaxy:center; Galaxy:kinematics and dynamics; Galaxy:structure}

\section{Introduction}
\label{sec:intro}

The presence of outliers in the velocity distribution of nearby stars
has been known for many years. Traditional  `runaway' stars, with
peculiar velocities typically greater than 40 km/s, constitute
approximately $30\%$ of O- and B-type stars in the
solar neighborhood \citep[e.g.,][]{Gies87, Stone91,Tetzlaff2011}.
Most runaway 
stars are ejected from the Galactic plane shortly after their formation
~\citep[][]{Ramspeck01}.

In 2005, a late B-type star was discovered in the outer halo
having a heliocentric radial velocity of 853 $\pm$ 12 km/s, which is
significantly in excess of the Galactic escape speed at that location
~\citep[][]{Brown05}. Such stars are called `hypervelocity stars'
\citep[HVSs; for a recent review see][]{Brown2015}.
\citet{Hills88} predicted the existence of HVSs, suggesting that they can
form via tidal disruption of a stellar binary by the central
supermassive black hole of the Milky Way. In this scenario, one of the
binary stars is captured by the black hole, while the other is
ejected, often at an extremely high velocity. To-date around 30
A/B/O-type unbound HVSs have been found \citep[see][]{Brown2015}. Such stars
can be used to probe star formation in the Galactic center or
constrain the potential of the Milky Way \citep[e.g.][]{Gnedin2005}.

The ejection rate of HVSs has been calculated in various studies and
has been used to investigate the properties of the progenitor
systems. For example, the work of \citet{Zhang13} consider models that
inject binaries from nuclear stellar disks adopting different stellar
mass functions.
They compare these to the observed S-stars in the Galactic Center and
the B-type HVSs in the Galactic halo. The models that best match the
set of observations have HVS ejection rates of $10^{-4}$ to $10^{-5}$
per year. Similar studies modeling the properties of HVS have
been carried out, for example \citet{Kenyon2014} or \citet{Rossi2014},
and even the possibility of an LMC origin has been considered
\citep{Boubert2016}.

Most other mechanisms are not expected to produce such high-velocity
stars, such as three-body interactions in the disk \citep{Poveda1967}
or super-nova kicks \citep{Blaauw1961}.
Recent works have made predictions for the ejection velocities from
these scenarios, showing that velocities in excess of a few hundred
$\rm km/s$ are unlikely \citep[e.g.][]{Portegies2010,Perets12}.
Some studies argue otherwise, for example \citet{Tauris15} has claimed
that asymmetric supernovae can produce
runaway stars with velocities as high as 1,000 km/s, but this work
cautions that more detailed population synthesis studies are required
to make robust predictions for the rates of such ejecta.

Although all currently known HVSs are young (O-/B-/A-type) stars,
given the slope of the initial mass function one may expect many more
older, low-mass HVSs \citep{Kollmeier2007}.
In an early attempt to find these, \citet{Kollmeier10} used the
non-detection of low-mass HVS in the Sloan Digital Sky Survey (SDSS)
to place an upper limit on their formation rate, concluding that the
ejection rate of unbound F/G-type stars is no more than 30 times
larger than that of B-type ejecta.

More-recently \citet{Palladino14} found 20 low mass G- and K-type HVSs
candidates from SDSS by incorporating proper motions, something which
\citeauthor{Kollmeier10} avoided. Half of these candidates exceeded
their corresponding escape velocity with at least $98\%$ probability,
but as none of the orbits were consistent with Galactic center ejection
their nature remained unclear. However, in the following year,
~\citet[][]{Ziegerer15} reanalyzed the proper motions for 14 of these
and found that they are all, in fact, bound to Galaxy and the initial
(incorrect) HVS classification was due to flawed proper motions.
Other low-mass hyper-velocity candidates have been claimed
\citep[e.g.][]{Li12}, but these may also be subject to spurious proper
motions.

Having realized the importance of reliable proper motions, 
\citet[][]{Vickers15} carried out a more cautious study of the runaway
population of low-mass stars in SDSS including more stringent proper
motion cuts. They were able to detect a number of high-velocity
runaway stars, but all of their hyper-velocity candidates were
marginal detections (with only one greater than 1$\sigma$). Therefore,
in effect, this result is in agreement with \citet{Kollmeier10}.

In this paper we continue the hunt for low-mass ejecta in SDSS,
focusing on high-quality proper motions and developing new techniques
for optimizing future searches. Note that the term HVS refers to stars
that have been ejected from an interaction with the supermassive black
hole {\it and} have velocities in excess of the escape speed at their present
location. However, since we are interested in all stars that have been
ejected by the Hills mechanism, regardless of whether or not they are
bound to the Milky Way, we follow \citet{Vickers15} and use the term
Hills star.

The outline of our paper is as follows. In Section \ref{sec:data} we
describe the SDSS Stripe 82 data used in our analysis. In
Section \ref{sec:candidates} we use SDSS spectroscopy to identify 226
Hills candidates, 29 of which are metal-rich.
In Section \ref{sec:palomar} we add our own data from
the Palomar telescope and identify one Hills candidate with [Fe/H]
$\sim$ -0.4 dex. In Section \ref{sec:halo} we discuss the possible
sources of contamination, most importantly halo stars on radial
orbits. We conclude in Section \ref{sec:conclusion}.

\section{Data and initial quality cuts}
\label{sec:data}
\subsection{Photometry}
\label{sec:photometry}

In order to identify Hills stars reliably, one of the most important
ingredients is a robust catalog of proper motions. Because of this, we
have chosen to focus our study on the Sloan Digital Sky Survey (SDSS)
Stripe 82 region. This is a narrow 2.5 deg wide stripe along
the celestial equator in the south Galactic cap, covering
250 deg$^2$ around $l \sim 60 $ to $ 190$ deg, $b \sim -20$ to $-60$ deg.
This stripe has been repeatedly imaged by SDSS over 7 years allowing
for studies of the variable sky and identification of many kinds of
transient phenomena ~\citep[e.g.][]{Sesar07, Watkins09}. This dataset
has also been used to construct high accuracy proper motion catalogs
~\citep[][]{Bramich08, Koposov13}. The ~\citet[][]{Bramich08} catalog
is complete down to $r\sim21.5$ mag, making it one of the deepest
large-area catalogs of its kind.
The catalog was subsequently revised in \citet{Koposov13}, removing
some low-level systematics and improving the uncertainties
by using quasars as a reference frame. The resulting catalog has a
precision of around 2 mas $\rm yr^{-1}$ and negligible systematic
errors ($\sim$0.1-0.2 mas $\rm yr^{-1}$).
These catalogs have been exploited to hunt for white dwarfs
~\citep[][]{Vidrih07} and wide binaries \citep{Quinn2009}, and analyze
the kinematic properties of Galactic disk and halo stars
~\citep[][]{Smith09b,Smith09a,Smith12}.

We also need good photometry, so that we can estimate distances using
the photometric relations from \citet{Ivezic08} and
\citet{Bond10}. Our photometry mainly comes from the \citet{Annis11}
catalog, where the multiple Stripe 82 epochs have been co-added to
result in a high precision catalog (e.g. $0.5\%$ photometry in
$g$,$r$,$i$ and $1\%$ in $u$). The $u$-band was sometimes problematic
and there are cases where no $u$-band magnitudes are reported; for
these stars we have used the standard single-epoch SDSS $u$-band
photometry \citep{Abazajian2009}, although we have not included any
stars with error greater than 0.2 mag.

We first de-redden our photometry using the maps of
\citet{Schlegel98}. To ensure good quality photometry and astrometry
we only use stars with magnitudes in the range 15 to 21 for all
$u,g,r,i$-bands. Second, we insist that the mean object
type $\geqslant 5.7$. This  requires an object to be classified as a star
in $\geqslant 90$ per cent of epochs. This is less than 100 per cent in
order to retain objects that have been misclassified in a limited
number of epochs due to problems with the SDSS star-galaxy separation
algorithm. In the \citet{Bramich08} catalog, this occurs
particularly in the final season, 
when observations were not exclusively taken in photometric
conditions. Third, we insist that the proper motion error is less than
4 $\rm mas$ $\rm yr^{-1}$ to clean any objects with problematic
astrometry. Then we identify objects on the stellar locus using the following
cuts \citep{Ivezic08}:
\begin{equation}
  0.7 < (u-g)_0 < 2.0,
\end{equation}
\begin{equation}
  -0.25 < (g-r)_0-5(u-g)_0  < 0.05,
\end{equation}
\begin{equation}
  -0.2 < 0.35(g-r)_0-(r-i)_0 < 0.1.
\end{equation}
where the subscript `0' denotes a color that has been extinction
corrected using the maps of~\citep[][]{Schlegel98}. 

Table \ref{tab:number} shows the number of stars passing these
photometric cuts. A total of 249,901 stars pass these criteria, of
which 36,869 use the SDSS single-epoch $u$-band magnitude (with the
rest having $u$-band magnitudes from the \citealt{Annis11}
co-add).

\subsection{Spectroscopy}
\label{sec:spectra}

To complement these proper motions, we need spectroscopy to
obtain radial velocities. We first obtain these from existing SDSS
spectroscopy, then later supplement this sample with our own
observations (see Section \ref{sec:palomar}).

The SDSS project, and in particular its Galactic extension SEGUE
\citep{Yanny2009}, have obtained spectra for hundreds of thousands of
stars. Stellar atmospheric parameters have been determined by the Sloan
Stellar Parameter Pipeline data product ~\citep[SSPP; see][]{Lee08},
and radial velocities found via ELODIE template matching.
We have cross-matched our photometric sample with the SDSS
DR10 spectroscopic catalogue, and found that $\sim$ 5.7\%
have suitable spectra -- defined as those with SSPP flag set to
`nnnn', which implies that there are 
no cautionary signs and the stellar parameters should be
well determined \citep{Lee2008}. In this work we limit ourselves to
spectra with $\rm S/N \ge 10$ and apply the following quality cuts:
$\rm \delta v_{helio} < 30 ~km/s$,
$\rm \delta log(g) < 0.5 ~dex$,
$\rm \delta [Fe/H] < 0.5 ~dex$.

Our analysis focuses on main-sequence stars, so we reject giant
contamination by removing all stars with
$\rm log(g) < 3 ~dex$\footnote{Even though this is not a very strong
  cut, it is unlikely our final sample contains giant contamination as
  our Hills candidates all have significant proper motions, which
  would not be the case for distant giant stars.}.
A total of 13,538 stars pass these cuts (see Table \ref{tab:number}).
The median uncertainties on the resulting sample are $\rm 2.1~km/s$,
$\rm 0.08~dex$ and $\rm 0.04~dex$, for $\rm \delta v_{helio}$,
$\rm \delta log(g)$, and $\rm \delta [Fe/H]$, respectively. These
uncertainties, which are those reported by the SSPP pipeline
\citep{Lee2008}, are estimates of the internal precision; if we fold
in external errors \citep{Allende08}, then the total uncertainties are
likely to be around $\rm 3.3~km/s$, $\rm 0.22~dex$ and $\rm 0.12~dex$,
for $\rm \delta v_{helio}$, $\rm \delta log(g)$, and $\rm \delta
[Fe/H]$, respectively.
For our calculations of the location and velocity of the last disk
crossing the main source of uncertainty comes from the proper motions,
and so the fact that these internal spectroscopic uncertainties may be
underestimated is of little consequence.

\begin{table*}
  \begin{center}
    \caption{The number of SDSS Hills candidates after the respective cuts.}
    \label{tab:number}
    \begin{tabular}{lllll}
      \hline
     Cut & Section & Description of cut & Number of objects\\
      \hline      
      1& \ref{sec:photometry} & Bright stars passing our photometric cuts & 249,901\\
      2 & \ref{sec:spectra} & High-quality SDSS stellar spectra & 14,174\\
      3 & \ref{sec:spectra} & Main-sequence stars ($\rm log(g) \ge 3 ~dex$) & 13,538\\
      4& \ref{sec:distance} & Color within range of the photometric parallax relation & 13,170\\
      5& \ref{sec:sdss} & SDSS Hills candidates & 226\\
      6& \ref{sec:sdss} & Metal-rich SDSS Hills candidates (SHCs) & 29\\
      \hline      
    \end{tabular}
  \end{center}
\end{table*}

\subsection{Distances}
\label{sec:distance}

To determine the full six-dimensional phase-space for each star, we
need to estimate distances. We do this using the relations presented
in \citet{Ivezic08}. These relations are valid for dwarf stars
in the color range $ 0.3\leqslant ( g - i)_0 \leqslant 4.0$, so we first reject
all stars outside this range (note that we remove giant stars by
enforcing $\rm log(g) < 3 ~dex$; see Section \ref{sec:spectra}). We
then estimate the distance using the color-magnitude relation given in
equations (A1), (A4) and (A5) of \citet{Ivezic08}. We also incorporate
the $\rm [Fe/H]$ dependence given in their equation (A2), where we use
the metallicity determined from the SDSS spectra.
\citet{Smith12} noted that the turn-off correction of \citet{Ivezic08}
is not ideal for general disk stars. They concluded that for the
magnitude range $0.3 < (g - i)_0 < 0.6$ a better choice is to have
no turn-off correction and to simply increase the uncertainty
\citep[equation A1 of][]{Smith12}. We adopt this approach for our
present study.

Combining the uncertainty in the \citet{Ivezic08} relation with the
error on $\rm [Fe/H]$, \citet{Smith12} found that for their Stripe 82
sample the median distance error was 10.7\%. Therefore in our current
work we simplify our calculation by assuming all stars have a distance
error of $10\%$. Our distances are mostly in the range 0.5 to 5 kpc,
although there are a small number of stars further out. This gives us
a final sample of 13,170 dwarfs with SDSS spectra (see Table
\ref{tab:number}).

\section{Identifying candidate Hills stars}
\label{sec:candidates}

\subsection{Calculation of the last disk crossing}
\label{sec:method}

Given the position and velocity of each star in our sample, we can
integrate the orbit back through time to calculate the last disk
crossing. Clearly Hills stars should have orbits whose last disk
crossing intersects the Galactic center.
We use potential 2b from \citet[][]{Dehnen98}, which is a
reasonable match to the more-recent model of \citet{McMillan11}.
\citeauthor{Dehnen98}'s model consists of three exponential disk components
(corresponding to the inter-stellar medium, the thin and thick disks),
together with a spherical bulge and halo. This is then fit to a
variety of observations in order to constrain the parameters of their
model.

Throughout this work we use a right-handed Cartesian coordinate system
centered on the Galactic Center (GC), with the x-axis pointing from the Sun
to the GC, the y-axis pointing in the direction of rotation and the
z-axis pointing towards the Northern Galactic Pole.
We assume that the motion of the local standard of rest (LSR) is
$\rm 238 \pm 9 ~km/s$, the velocity of the Sun with respect to the
LSR is $\rm (14.0 \pm 1.5, ~12.24 \pm 0.47, ~7.5 \pm 0.3)~km/s$, and the
Sun lies at $\rm x = -8.27 \pm 0.29$ kpc \citep{Schonrich12}.

Orbits are calculated for every star in our sample by reversing
their velocities and running them through potential 2b for 1
Gyr with a time-step of 1 Myr resolution. The choice of a 1 Gyr
integration time is to ensure we find crossings where possible.
When an orbit crosses the Galactic plane, a linear interpolation
between the calculation steps before and after the crossing is used to
find the exact coordinates and velocities of the crossing point. We
assume here that the most-recent crossing time was the point where 
the object was ejected, if it was ejected. We 
refer to this most-recent crossing location as the `crossing position'
($\rm {\bf r_{cr}}=[x_{cr}, ~y_{cr}, ~0]$),
and the speed at this point as the `crossing velocity' ($\rm |{\bf v_{cr}}|$).

Uncertainties in the crossing position are calculated using a Monte
Carlo approach. We assume all observational errors are Gaussian and
generate 400 realizations for each star in our sample. We account for
all sources of uncertainty, namely the helio-centric distance, radial
velocity, proper motion, together with uncertainties in the LSR and
solar radius.

For each star we use the distribution of ($\rm x_{cr}, ~y_{cr}$) from
our 400 realizations to estimate the crossing radius and uncertainty
as follows,
{\small
\begin{eqnarray}
\rm r_{cr} &=& \vert \left< {\bf r_{cr}} \right>\vert = \rm \sqrt{ \left< x_{cr} \right> ^2 + \left<
  y_{cr} \right> ^2}\label{eq:rej}\\
\rm \delta r_{cr} &=& \rm \frac{\sqrt{\left<x_{cr}\right>^2 \sigma
    \left(x_{cr}\right)^2 + \left<y_{cr}\right>^2
    \sigma\left(y_{cr}\right)^2 + \left<x_{cr}\right>\left<y_{cr}\right>
    \sigma\left(x_{cr} , y_{cr}\right) }}{r_{cr}}\nonumber\\
\end{eqnarray}}
where angle brackets denote the mean of the realizations, $\sigma(X)$
denotes the standard deviation and $\sigma(X,Y)$ denotes the square
root of the covariance. Note that since the distribution of
$\rm r_{cr}$ values does not follow a normal distribution, it would be
incorrect to take the standard deviation to represent the uncertainty.
The correct uncertainty on this quantity can be obtained through error
propagation, as given in the above equation.

For the crossing velocity we take the value corresponding to 
the crossing radius calculated from equation (\ref{eq:rej}). We do
this, rather than use the mean from our realizations, because
$\rm v_{cr}$ and $\rm r_{cr}$ are strongly 
correlated. One consequence of this approach is that if there are
any Hills stars in our sample, this value of $\rm v_{cr}$ will
significantly underestimate the true ejection velocity; even if the
probability distribution function of
$\rm r_{cr}$ indicates that it is consistent with zero, our adopted
value of $\rm r_{cr}$ can never be precisely zero (due to an
offset from observational errors), and hence the corresponding 
$\rm v_{cr}$ will be lower than the ejection velocity corresponding to
$\rm r_{cr}=0$.

Once we have identified our sample of Hills candidates, we repeat this
procedure at higher resolution (using 5,000 realizations) in order to
improve our estimates. Throughout this paper we calculate
probabilities of crossing location using the actual distributions of
our realizations; the above errors are only quoted in the table of
candidates.

\subsection{Candidates from SDSS spectroscopy}
\label{sec:sdss}

\begin{figure}
\plotone{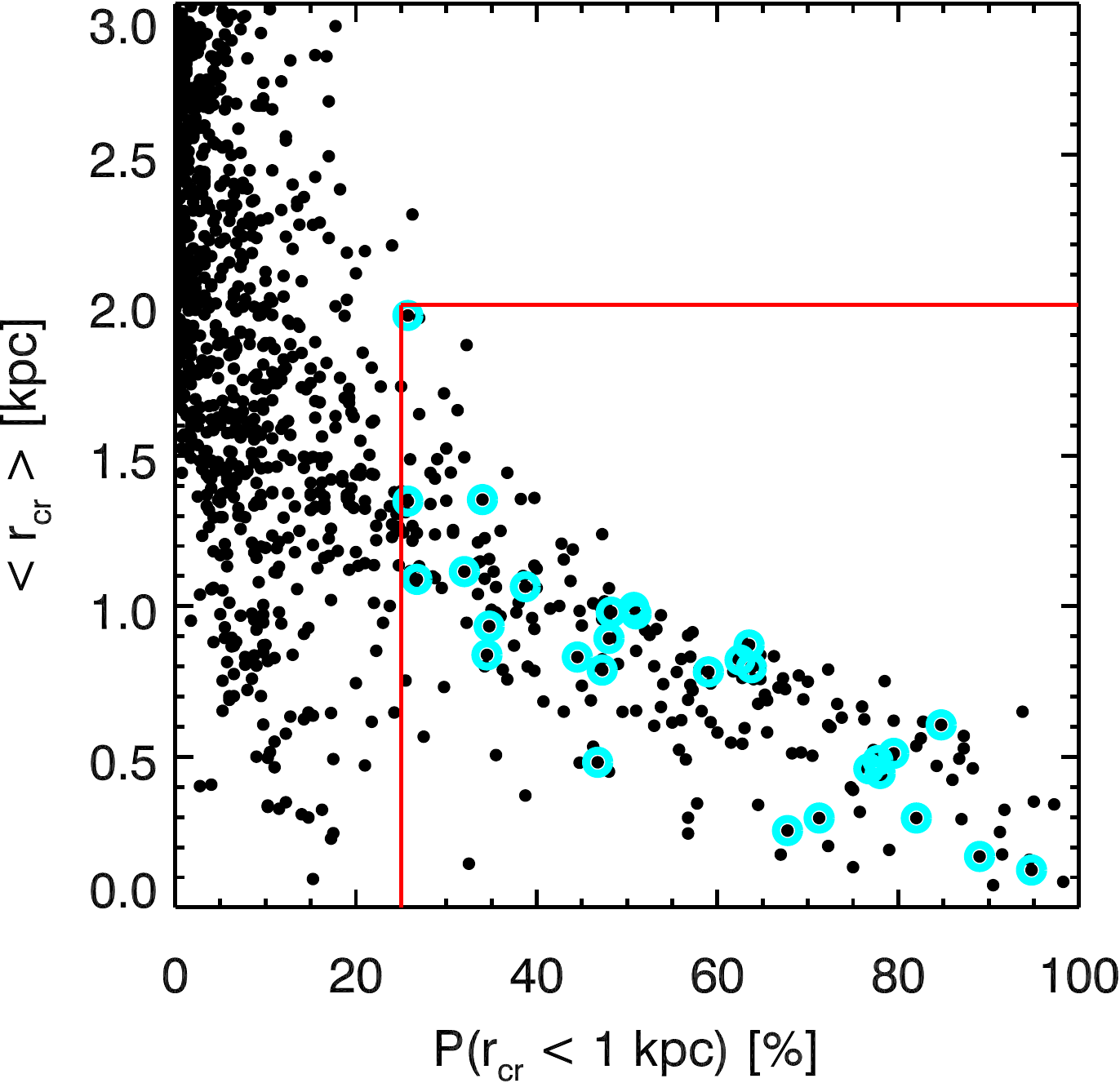}
\caption{This figure shows the mean value of the last disk crossing
  radius $\rm \left(r_{cr}\right)$ versus the probability that the
  last disk crossing radius is less than 1 kpc for all main-sequence
  stars with good SDSS spectroscopy and distances in Stripe 82 (Cut 4
  from Table \ref{tab:number}). Cyan points show our 29 metal-rich
  Hills candidates and the red line shows our selection boundary.}
\label{fig:cuts}
\end{figure}

We identify good candidates using the probability distributions for
$\rm r_{cr}$ from the Monte Carlo resamples.
We first obtain an initial sample of Hills candidates by rejecting all
stars with $\rm \bf\vert \left< r_{cr} \right> \vert $ greater than 2 kpc and only
retaining stars with $\rm P \left (r_{cr} < 1 kpc \right) >
25 \%$. These two cuts are illustrated in Fig. \ref{fig:cuts}.
This latter cut is important as it removes stars with crossing
positions that are highly uncertain or unlikely to be coincident with
the Galactic centre.

We do not expect all of these to be genuine Hills stars, with the
main source of contamination being halo stars on radial orbits. To
remove halo contamination we impose a cut on the metallicity,
requiring $ \rm [Fe/H] > -0.8$ dex. Note that the metallicity
distribution of the halo is well described by a normal distribution
with mean $-1.46$ dex and dispersion $0.3$ dex
\citep{Ivezic08,Bond10}, while bona-fide Hills stars are expected
to have been born in the central parts of the galaxy and should
therefore be metal-rich. This cut should remove most halo contamination,
but we leave a detailed discussion of halo contamination to Section
\ref{sec:halo}.

\begin{figure*}
    \plotone{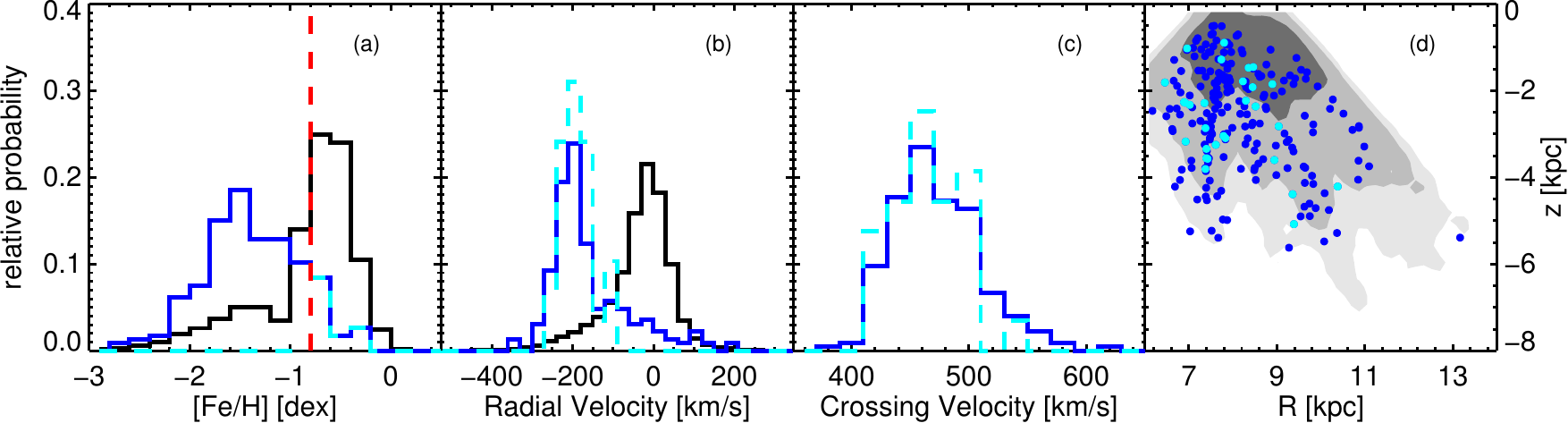}
    \caption{The various properties of our Hills samples: (a) the
      metallicity, (b) radial velocity, (c) velocity at last disc
      crossing and (d) current spatial distribution in Galactocentric
      cylindrical coordinates.
      Black lines correspond to the entire sample of SDSS dwarfs (Cut
      4 from Table \ref{tab:number}), the blue lines correspond to the
      full sample of 226 Hills candidates (Cut 5) and the cyan lines
      to the 29 metal-rich candidates ($\rm [Fe/H] > -0.8$ dex; Cut 6).
      In panel (d) we use contours to denote the distribution of the
      full spectroscopic sample, using dark grey for 1$\sigma$, grey
      for 2$\sigma$, and light grey for 3$\sigma$.}
    \label{fig:rb}
\end{figure*}

A total of 29 Hills candidates pass this metallicity cut, and we label
these SDSS Hills Candidates (SHC1--29). The properties of these 29
stars are shown in Figure \ref{fig:rb}, alongside our larger sample of
226 candidates and the full spectroscopic sample of 13,170 dwarfs (see
Table \ref{tab:number}).
Clearly most stars in the full sample are regular disk stars, with
metallicities around -0.5 dex and radial velocities around 0 $\rm km/s$.
However, most of 226 Hills candidates are metal-poor, confirming our
suspicion that halo stars are an important source of contamination.
The far-right panel of Fig. \ref{fig:rb} shows the current spatial
distribution of these samples, where because of the orientation of
Stripe 82 these objects are below the plane and are at Galacto-centric
radii of between 6 and 10 kpc. The distribution is fairly homogeneous
across the Stripe 82 field.
The center-left panel shows the radial velocity distribution. The
Hills candidates have high radial velocities, but not exceptionally
high, i.e. they are similar to other high velocity stars in the solar
neighborhood. This means that one cannot identify such Hills stars
using radial velocity alone, unlike the previously detected
unbound hyper-velocity stars \citep[e.g.][]{Brown14}, and all of our
candidates are bound to the Milky Way potential.
Because of the orientation of the orbits in our sample, most of these
Hills candidates have negative radial velocities, i.e. are moving
towards the Sun.
The center-right panel shows the distribution of crossing
velocities\footnote{As explained in Section \ref{sec:method}, this
is the velocity corresponding to the crossing radius calculated from
equation (\ref{eq:rej}) and, if these are bona-fide Hill stars, then
their true ejection velocities would be significantly greater.}.
The crossing velocities of our Hills candidates are mostly around 450
$\rm km/s$ or more, which is necessary for them to make the journey
from the central parts of the Galaxy to the Solar neighborhood.
As discussed in Section \ref{sec:intro}, such ejection speeds are
unlikely to be obtained through other mechanisms.
\citet{Perets12} showed that dynamical (three-body)
encounters in clusters are unlikely to produce ejection velocities of
hundreds of $\rm km/s$, especially for low-mass stars (see Fig. 3 of
their paper). The likelihood of this occurring from a supernovae binary
ejection is also negligible; velocities of up to 200 or 300 $\rm km/s$ are
possible, but unlikely for low-mass stars like the ones we consider
here \citep[e.g. Fig. 6 of][]{Portegies2010}.

Table \ref{tab:hills_metalrich} lists the properties of our 29 metal-rich Hills
candidates. The proper motions and $g$-band magnitudes are obtained
from SDSS Stripe 82 data \citep{Koposov13,Annis11}. The metallicities and
heliocentric radial velocities come from SDSS spectroscopy. As can
be seen from their total velocities, all are bound to the Galaxy,
assuming the escape speed is 500--600 $\rm km/s$ \citep{Smith07,Piffl14}.  

\begin{figure*}
\begin{center}
\includegraphics[width=15cm]{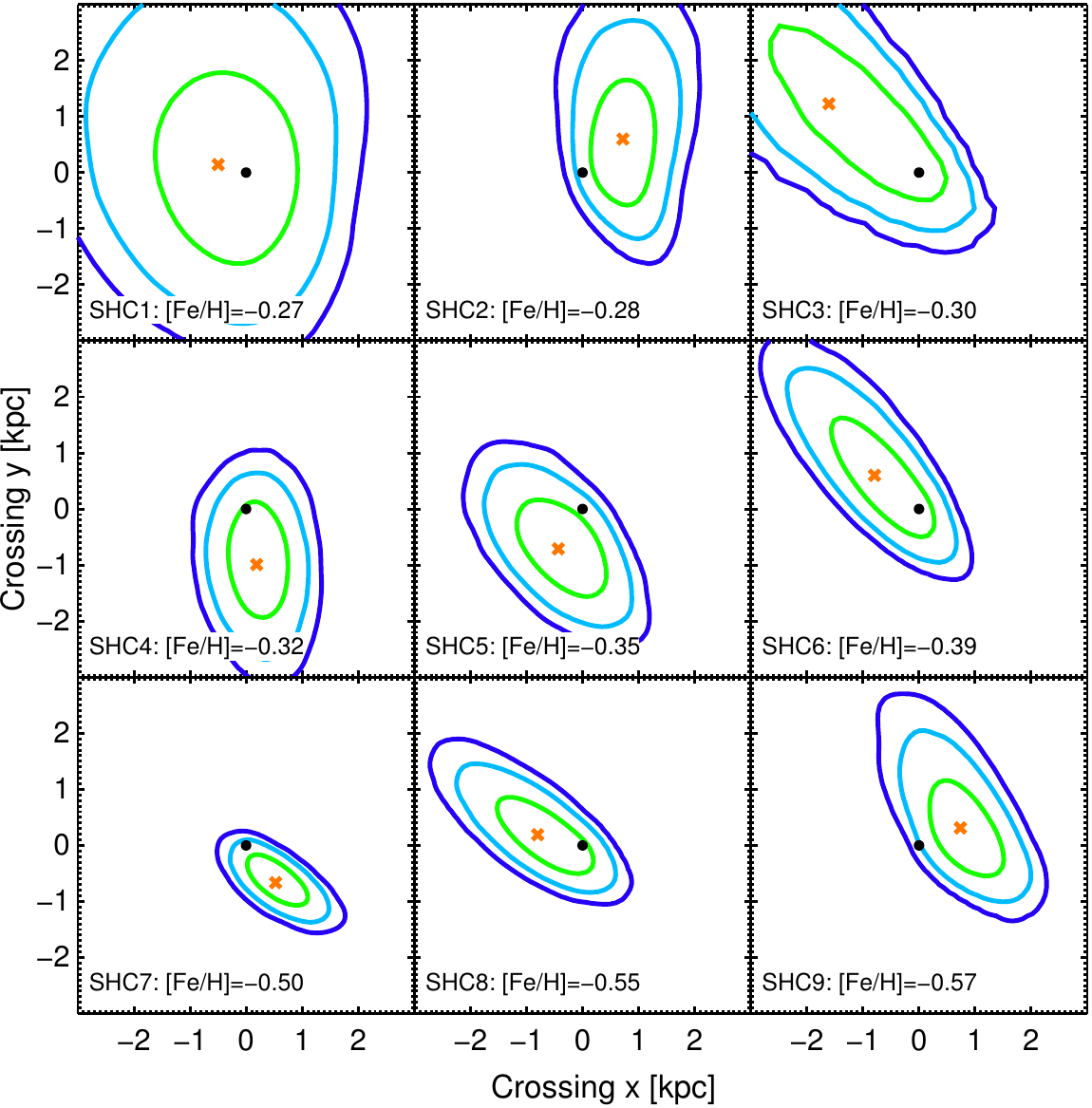}
  \caption{This figure shows the last disk crossing position
    for our nine most metal-rich SDSS Hills
    candidates (i.e. with $[Fe/H] \geqslant -0.57$ dex).
    The green, cyan and blue lines correspond to
    1$\sigma$, 2$\sigma$ and 3$\sigma$, respectively. The orange
    crosses show the crossing position if we do not account for
    observational errors. The black dot shows the position of Galactic
    center. The remaining candidates are shown in Fig. \ref{fig:hills_candidates}.}
    \label{fig:hills_metalrich}
\end{center}
\end{figure*}

The first ten rows of Table \ref{tab:hills_metalrich} show our best Hills
candidates, in that their metallicities are all greater than $-0.6$
dex (and hence less-likely to be halo stars; see Section
\ref{sec:halo} for further discussion on this point). Their typical
last crossing velocities are larger than 400 km/s and for most the
last crossing radii are within a kpc of the Galactic center. In Figure
\ref{fig:hills_metalrich}, we show the probability distribution for
the crossing position of the nine most metal-rich Hills
candidates. The uncertainties come from our Monte Carlo calculation
(see Section \ref{sec:method}).
For five of these nine stars (SHC1, SHC3, SHC4, SHC6 and SHC8) the
Galactic center is within $1\sigma$ of the crossing position and hence
they are especially deserving of further study.
Amongst our larger sample of 29 metal-rich SDSS Hills candidates, the
GC lies within $1\sigma$ of the crossing position for around half of
all candidates. The distributions of last disk crossing location for
these candidates are presented in the Appendix.

\begin{table*}
 \begin{center}
   \caption{The properties of our metal-rich ($\rm [Fe/H] > -0.8$)
     Hills candidates.}
   \label{tab:hills_metalrich}
   \begin{tabular}{lccccccccccccccc}
     \hline
     name &IAU name& $\mu$ & g &[Fe/H]$^a$& d$_{helio}^a$ &v$_{helio}^a$
     &$\rm v_{tot}$& $\rm v_{cr}$ & $\rm r_{cr}$&P($\rm r_{cr}<1$kpc)\\ 
     & & (mas/yr) & (mag)&dex&kpc&km s$^{-1}$&km s$^{-1}$&km s$^{-1}$&kpc&\%\\

     \hline
SHC1 &J014305.79-003903.8  &  10.33 $\pm$ 1.09  & 18.53 &  -0.27 &   4.83 & -104.8 $\pm$ 
3.9 &  258.7  & 509.8  &  0.50 $\pm$ 0.86 & 47.8   \\
SHC2& J004614.45-010346.9 &  16.13 $\pm$ 1.09  & 18.40  & -0.28  &  3.14 & -152.0 $\pm$
2.6 &  284.4  & 450.0 &  0.94 $\pm$ 0.36 & 50.0    \\ 
SHC3&J221807.24+005439.8 &   9.12 $\pm$ 1.16   & 19.02  & -0.30  &  5.17 & -229.1 $\pm$
3.9 &  320.1  &410.4 & 2.05 $\pm$ 1.57 & 24.9    \\ 
SHC4& J001740.99-002759.1  &  10.03 $\pm$ 0.83  & 18.28  & -0.32  &  4.07 &  -98.6 $\pm$
3.2  & 217.3   & 459.5&  1.00 $\pm$ 0.36 & 46.8   \\ 
SHC5&J213915.13+011006.1 &   8.84 $\pm$ 1.53   & 18.30  & -0.35  &  3.98 & -150.6 $\pm$
2.2 &  224.6  & 489.9 &  0.82 $\pm$ 0.32 & 49.4    \\ 
SHC6& J220317.08+010539.7 &   9.98 $\pm$ 1.69   & 17.95  & -0.39  &  3.51 & -195.1 $\pm$
2.1  & 256.2  & 427.8  &  0.98 $\pm$ 0.82 & 51.6   \\ 
SHC7& J221331.35+005508.2&  14.73 $\pm$ 0.74  & 16.79  & -0.50  &  1.89 & -207.5 $\pm$
1.1 &  245.8  & 440.8  &  0.84 $\pm$ 0.41 & 65.9  \\ 
SHC8 &J220347.45-010526.0  &  29.28 $\pm$ 0.94  & 17.14  & -0.55  &  1.34 & -196.6 $\pm$
2.0 &  270.3  & 442.1  &  0.82 $\pm$ 0.62 & 62.1   \\ 
SHC9 &J223257.10+002939.5 &   7.69 $\pm$ 0.89   & 18.65  & -0.57  &  4.46 &  -246.5 $\pm$
5.1 & 295.3 &  455.0 & 0.80 $\pm$ 0.25 & 55.0   \\ 
SHC10& J221952.86-011254.2 &   8.29 $\pm$ 0.99   & 18.67  & -0.59  &  4.03 & -173.3 $\pm$
4.1  & 234.6  &421.2 &  1.12 $\pm$ 0.41 & 25.3  \\ 
SHC11& J221738.85+003322.3 &   6.09 $\pm$ 1.49  & 19.18   &-0.61   & 4.83  &-244.8 $\pm$
3.5   & 281.7  &431.0  &  1.43 $\pm$ 0.91 & 29.3   \\
SHC12& J002458.60-010832.5 &   7.41 $\pm$ 1.11  & 18.08   &-0.64  &  5.69  &-142.6 $\pm$
6.9  & 245.4   & 476.3 &  0.81 $\pm$ 0.76 & 35.7    \\ 
SHC13&   J005845.36+005346.9   & 18.60 $\pm$ 1.02   & 17.61   &-0.65  &  2.10  &-159.3 $\pm$
  1.8  & 244.2   & 451.0 &  0.94 $\pm$ 0.20 & 34.9    \\ 
SHC14& J221711.45+001937.6   &   7.73 $\pm$ 1.44  & 19.53   &-0.66  &  5.49  &-231.4 $\pm$
7.5  & 306.7  & 459.5   & 0.83 $\pm$ 1.22 & 46.5   \\ 
SHC15& J222229.85-002903.3   &   7.10 $\pm$ 1.43  & 19.68   &-0.68  &  4.99  &-224.8 $\pm$
5.5  & 280.7  & 499.2   & 0.27 $\pm$ 0.35 & 69.5  \\ 
SHC16&  J205250.90+003453.6   &  6.32 $\pm$ 1.64   & 18.52   &-0.68  &  4.08  &-233.0 $\pm$
 4.0  & 263.0  &495.1 &  0.51 $\pm$ 0.57 & 78.7  \\ 
SHC17& J235508.19+005619.5 &  17.71 $\pm$ 0.94 &  16.83  & -0.69 &   2.24 & -197.4 $\pm$
2.1  & 272.9  & 432.7&   1.09 $\pm$ 0.68 & 39.6   \\ 
SHC18& J213412.37-004311.0 &   8.08 $\pm$ 1.04  & 18.09   &-0.70  &  3.87  &-204.3 $\pm$
2.2  & 252.3  &451.9 & 0.51 $\pm$ 0.58 & 77.9    \\ 
SHC19& J214714.91-000227.3  &   5.23 $\pm$ 1.73  & 19.32  &-0.71  &  5.14  &-198.4 $\pm$
6.1  & 235.7 &  474.0 & 0.27 $\pm$ 0.72 & 67.1  \\ 
SHC20& J001327.34-010136.5 &  26.97 $\pm$ 0.64 &  16.40  & -0.72 &   1.65 & -217.9 $\pm$
2.0 &  303.0 &  497.9 & 0.43 $\pm$ 0.76 & 76.8  \\ 
SHC21& J232728.65+005409.6 &  20.35 $\pm$ 0.83 &  17.95  & -0.72  &  2.17 & -158.1 $\pm$
2.1 &  262.3  & 472.3 & 0.59 $\pm$ 0.28 & 86.3  \\ 
SHC22&  J003525.74-010942.1 &   9.48 $\pm$ 1.12 &  18.87   &-0.74  &  4.89  &-161.1 $\pm$
5.9  & 272.7  & 441.0 & 1.45 $\pm$ 1.20 & 24.3   \\ 
SHC23&  J225448.69-005430.2  &   7.75 $\pm$ 0.96 &  19.05   &-0.76  &  3.95  &-182.1 $\pm$
2.2  & 232.8  & 452.9  & 0.91 $\pm$ 0.37 & 43.5   \\ 
SHC24& J213137.16+005736.7   &   6.78 $\pm$ 1.65 & 19.27   &-0.76   & 4.04 & -225.9 $\pm$  
6.1  & 260.5  & 472.4 & 0.46 $\pm$ 0.09 & 76.0   \\ 
SHC25& J234839.28+010320.3 &  25.04 $\pm$ 1.73 & 17.40  & -0.76  &  1.74 & -153.5 $\pm$
2.2 &  257.2 &  503.4 & 0.16 $\pm$ 0.07 & 91.2  \\ 
SHC26& J204555.46-004653.3   &  9.13 $\pm$ 1.66  & 17.04   &-0.77   & 2.38  &-222.9 $\pm$
 2.0  & 245.5  & 427.5 & 1.09 $\pm$ 0.47 & 33.7    \\ 
SHC27& J233657.12-002138.7 &  14.77 $\pm$ 0.90 &  17.43 &  -0.77  &  2.64 &  -209.0  $\pm$
1.8 &  279.0 &  490.5 & 0.33 $\pm$ 0.56 & 80.2   \\ 
SHC28&  J224722.50+000802.0 &  10.93 $\pm$ 0.93  & 18.52 &  -0.78 &   4.00 & -187.1 $\pm$
2.2  & 279.1  & 451.8 & 0.83 $\pm$ 0.22 & 61.5  \\ 
SHC29&  J235446.02+010819.4   &  19.53 $\pm$ 0.68  & 17.09  & -0.80 &   2.77 & -107.5 $\pm$
2.6 &  278.2  & 540.2 & 0.11 $\pm$ 0.06 & 94.6  \\ 
\hline
PHC1&J225957.20+000543.8  &  $8.20\pm1.07$ & 18.59 & -0.40 &  4.10
&$-124.4\pm10.0$ &202.2& 393.89 &
1.70$\pm$0.46&3.22\\ 

PHC2 &J225826.60+005013.6 &  $13.44\pm0.72$ &  18.20 &  -0.33
&  3.30& $-79.1\pm2.4$ &224.6&  358.62 & 2.58$\pm$0.34&0.00\\ 

PHC3 &J231828.40-004040.6 &   $18.02\pm1.06$ &  17.74 &-0.23 & 2.72&
$-203.3\pm5.9$ &308.8& 430.91 & 1.74$\pm$1.71&25.41\\

PHC4  &J214056.00+010020.0 &  $22.65\pm1.15$ &  18.20 &  -0.22 &
2.82&$-194.0\pm5.0 $&358.7&510.16 & 1.34$\pm$0.42&11.85\\ 

PHC5   &J222624.50-011416.0  &  $8.54\pm0.90$  & 18.40 &  -0.22&
4.48 & $-70.4\pm 2.4$&194.6&357.13 &
2.84$\pm$0.45&0.02\\ 
     \hline
   \end{tabular}
 \end{center}
 {\bf Notes:} SHC1--26 are candidates with SDSS spectroscopy (Sections
 \ref{sec:data} and \ref{sec:candidates}), while
 PHC1--5 have P200 spectroscopy (Section \ref{sec:palomar}).
 The typical distance errors are around 10-15\%. Metallicity errors
 are around 0.1 dex for the SHC candidates and 0.2 dex for the PHC
 candidates. $r_{cr}$ and $v_{cr}$ denote the position and velocity
 at the last disk crossing (see Section \ref{sec:method}).
\end{table*}

\section{Optimizing candidate selection from proper motion surveys}
\label{sec:palomar}

The efficiency of detecting Hills candidates from a blind spectroscopy
survey (i.e. one with no preferential targeting of Hills candidates)
is clearly limited. The SDSS spectroscopic catalogue described
above had an efficiency of only around 0.1\%, with 10-20 Hills
candidates coming from a total of 13,170 stellar spectra in
the Stripe 82 region. In this section we describe an approach to
optimize the search efficiency by utilizing proper motion information,
then apply this technique to obtain candidates using spectroscopy from
the Palomar 200-inch telescope.

\subsection{Target selection}
\label{sec:target_selection}

We again base our target selection on the high-quality Stripe 82
proper motions \citep{Bramich08,Koposov13}. Targets are selected with
SDSS $g$-band magnitudes between $16 < g < 19$, which balances our
desire to observe many candidates with the need for sufficient $S/N$
to measure reliable line-of-sight velocities and metallicities.

As can be seen in the previous section (e.g. the top-left panel of
Figure~\ref{fig:rb}), most of our Hills candidates are metal-poor
($\rm [Fe/H]\lesssim-1.0$), but stars that have been ejected from the
Galactic center are expected to be relatively metal-rich. We thus use
photometric metallicity relations to focus our search on metal-rich
stars, utilizing the relations presented in \citet{Ivezic08} and
\citet{Bond10}.
This method is calibrated for F- and G-type main-sequence stars, which
we select by applying a color cut of $0.2\leqslant (g - r)_0 \leqslant 0.6$ to
the Stripe~82 catalog; this is in addition to the cuts described in
Section \ref{sec:photometry} and Section \ref{sec:distance} that are
imposed to deliver stars on the stellar locus with good photometry.
We choose to reject all stars with photometric metallicity $\rm [Fe/H]
< -0.6$, leaving a total of 21,052 stars that pass all of these
cuts.

Distances are calculated using the approach described in
Section \ref{sec:distance}.
Because the errors on individual photometric metallicities are larger
than spectroscopic measurements, the errors in our derived distances
are correspondingly larger at around 20\%. We use these distances to
estimate the current 3D positions and tangential velocities of all
stars satisfying the above cuts, using the same approach as described
in Section \ref{sec:distance}.

To select the best candidates for Palomar observations, we simulate
the orbits of all 21,052 stars  for a range of radial velocities.
In the top-right panel of Figure~\ref{fig:rb}, it can be seen that
most of the radial velocities for our Hills candidates are
negative. If we calculate the last disk crossing (i.e. potential
ejection position) for each star as a function of the unknown radial
velocity and focus on stars with minimum crossing position less than
0.5 kpc, we find that the minimum crossing position almost always
occurs when the radial velocity lies in the range $-450$ to $-50$ \kms
(only 1\% of these stars have velocities outside this range). We use
this velocity range in the following analysis.

\begin{figure}
  \plotone{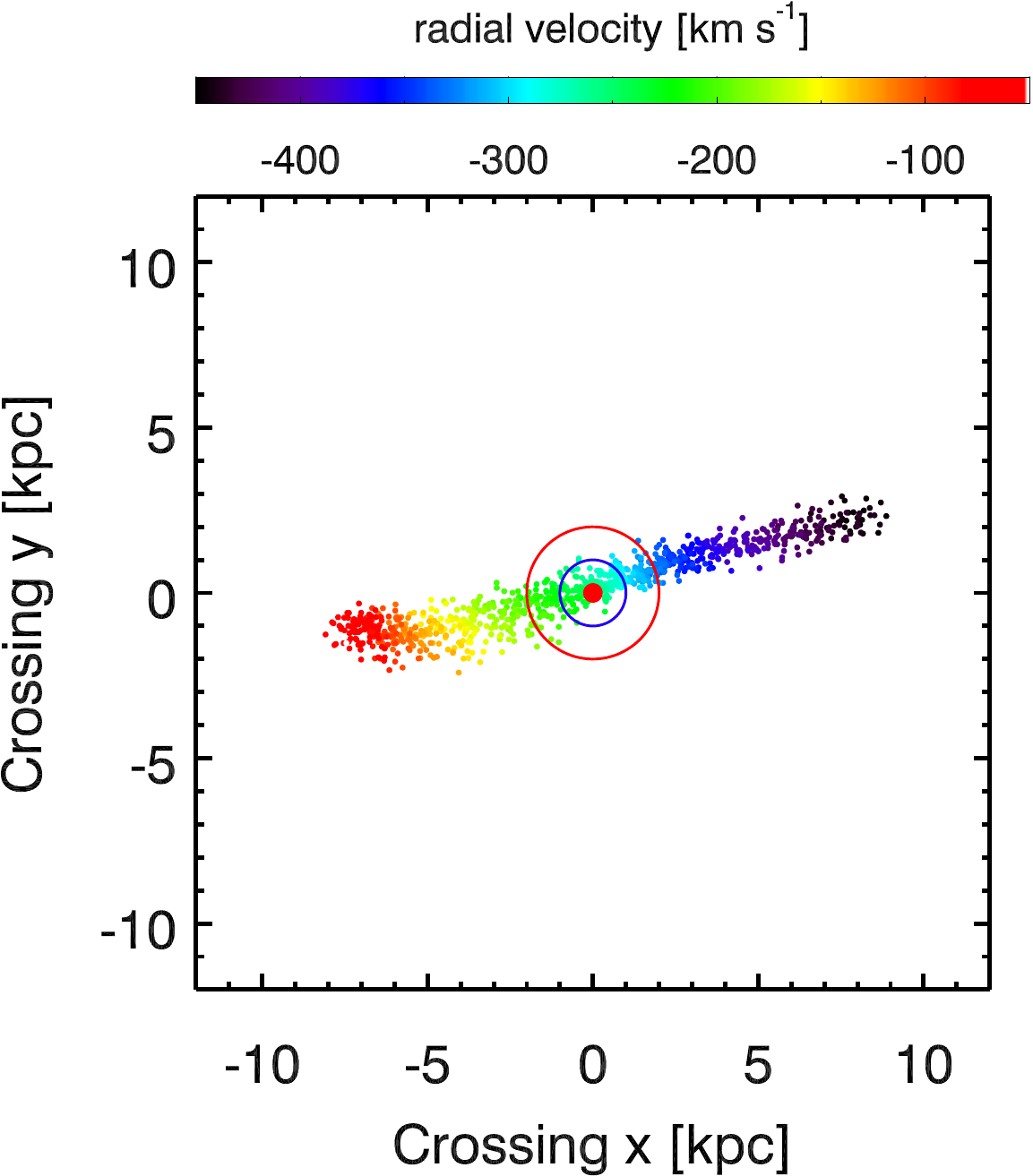}
  \caption{An example of a crossing track, showing how the
    location of the last disk crossing varies as a function of the
    radial velocity. Tracks which intersect the Galactic centre are
    candidate Hills ejecta. The width of the track is determined by the
    observational errors (mainly in the distance and proper motion)
    and the colors correspond to the radial velocity. The red point
    corresponds to the Galactic centre, and the blue and red circles
    denote 1.0 and 2.0 kpc radii, respectively.}
\label{fig:vrad}
\end{figure}

For each of our 21,052 stars, we calculate 400 Monte Carlo
realizations of the orbit. We do this following the approach described
in Section \ref{sec:method}, incorporating errors on all observable
quantities (helio-centric distance, proper motion, LSR and solar
radius), but since we do not know the radial velocity we choose this
uniformly within the range [$-450$, $-50$] \kms. An example is
illustrated in Figure \ref{fig:vrad}, where each point corresponds to
the last disk crossing location for an individual Monte Carlo
realization of the orbit\footnote{
Note that for the sake of clarity this figure uses 1000
realizations. Our candidates were selected using only 400
realizations in order to minimize computational time.}. Potential
Hills candidates can be identified using this crossing track,
with good candidates being those for which the track intersects the
Galactic centre (such as this example).

We classify the strength of these candidates using two quantities
derived from these Monte Carlo simulations:
$N_{1.0}$ is the number of realizations for which the last disk crossing
position lies within 1 kpc of the GC, and $N_{0.5}$ is number of
realizations within 0.5 kpc of the GC.
We define two sets of candidates, where the second is included to 
provide additional candidates at high Right Ascension,
compensating for the fact that most candidates are at low Right
Ascension and would not be observable for the whole night during
our run.

\begin{itemize}

\item Main targets: $N_{1.0} >$ 124 ($\sim$30\%), $N_{0.5} >$ 58 ($\sim$15\%), $\rm [Fe/H] >-0.45$;
\item High-RA targets: $\rm RA > -30^\circ $, $N_{1.0} >$ 99 ($\sim$
  25\%), $N_{0.5} >$ 39 ($\sim$10\%), $\rm [Fe/H] >-0.6$.

\end{itemize}

We select 60 main and 16 high-RA targets. Two good targets were later
rejected because they lie close to bright stars and would be difficult
to observe.
 
\subsection{Palomar 200-inch observations}
\label{sec:pal_obs}

Observations were carried out on the nights of 29-31 August 2014 using
the Double Spectrograph \citep[DBSP;][]{Oke82} on the Hale 200-inch (P200),
with time awarded from China's Telescope Access Program. DBSP
splits the light by use of a dichroic, allowing simultaneous
observation of blue and red channel spectra. For our purposes, we used
the D55 dichroic to place the transition at $\sim5500$~\AA. All
observations used a 128$''$-long slit with a width of 1.0~arcsec, with
typical seeing on all three nights of $\sim0.8-1.2''$. In the blue
channel, we used the 600 line/mm grating (600/4000) centered 
at $\sim$~4500~\AA, and the 600 line/mm grating (600/10000) centered at
$\sim$~7200~\AA~in the red channel. This set up achieved resolutions
of R$\equiv \frac{\lambda}{\Delta\lambda} \sim$1800 and R$\sim$2400 at
the positions of the hydrogen Balmer (H$\beta$ and H$\alpha$) lines in
the blue/red channels, respectively, with nearly continuous wavelength
coverage from $\sim$ 3000 - 8800~\AA. Each star was observed in two
exposures, with typical exposure times ranging from $300-600$~seconds,
depending on the target magnitude.

DBSP separates spectra into two channels. With our setup, the blue
band covers $\sim3000-6020$~\AA, and the red band $\sim5520 -
8870$~\AA. To reduce the targets spectra, we use the standard tasks
in IRAF. We observed calibration lamps at the beginning of each night,
including a HeNeAr lamp for the red channel and FeAr to fix the blue
channel wavelength solution. However, after reducing the spectra, a
check of sky lines with known wavelengths (including those at
4046.6~\AA, 4358.3~\AA, 5461.0~\AA, and 5577.3~\AA~in the blue band,
and 7750.6~\AA, 7821.5~\AA, 7913.7~\AA, 7993.3~\AA, 8344.6~\AA,
8382.4~\AA, 8399.1~\AA, 8430.1~\AA, 8452.3~\AA, and 8493.4~\AA~in the
red channel) shows offsets from their expected wavelengths. We thus
correct each spectrum by this offset to place the sky lines at zero
radial velocity.

For late-type stars, we predominantly use the calcium (Ca
\uppercase\expandafter{\romannumeral2}) H and K 
lines (3933.7~\AA~and 3968.5~\AA) to compute the average
wavelength shift and obtain the best-fitting radial velocity in the
blue band. In the red band, we use the near-infrared Ca
\uppercase\expandafter{\romannumeral2} triplet at 8498.0~\AA,
8542.1~\AA, and 8662.14~\AA~to compute the average wavelength shift
and obtain the best-fitting radial velocity. The mean errors for all
spectra are less than 20 km~$\rm s^{-1}$ in the blue band and less
than 10 km~$\rm s^{-1}$ in the red. We choose to adopt the red-band
velocities because of the higher resolution and the fact that there
are more sky lines in this region. For the fainter stars we took two
separate exposures in the red band and use the average of the two
velocities as our final value. The two separate measurements show
good agreement within the expected observational errors.

As discussed above, metallicities are important for determining the
likelihood that a star was ejected from the GC. We estimate
metallicities using a method based on empirical fits to the behavior
of [Fe/H] as a function of spectral line indices. In particular, we
measure the equivalent widths of spectral lines using bandpass and
pseudo-continuum definitions defined by \citet{Worthey94} for the
low-resolution, Lick/IDS absorption line indices \citep{Burstein84,
Faber85, Gorgas93}. The set of standards defining our empirical
calibration include the standard system of \citet{Schiavon07}, and the
MILES \citep{SanchezBlazquez06}, the ELODIE.3.1 \citep{Moultaka04},
and STELIB \citep{LeBorgne03} libraries of spectra. We derive surface
fits to the behavior of [Fe/H] as a function of the Fe5270
vs. H$\beta$ indices (which are primarily sensitive to iron and
hydrogen abundances). In practice, we average a number of Fe indices
from the spectra to derive an average iron index. The code recovers
[Fe/H] from high $S/N$ spectra of stars with known metallicities with
accuracy of $\leqslant0.3$~dex. This method has been used on low-resolution
spectra of stars in the Sagittarius stream \citep{Carlin12}, the
Bo\"{o}tes~III dwarf galaxy \citep{Carlin09}, and the Anticenter
Stream \citep{Carlin10}, among others. We derive metallicities by
running this algorithm on each of the spectra we obtained with
DBSP and again adopt the values from the red part of the spectra. In
general the agreement with the photometric metallicities is
reasonable, with (as expected) a spread of around 0.4 dex. There are a
couple of stars with spectroscopic values that are significantly more
metal-poor than the photometric estimate, but as we describe below
these are removed from our analysis.

\subsection{Crossing positions and evaluation of success rate}

\begin{figure}
\plotone{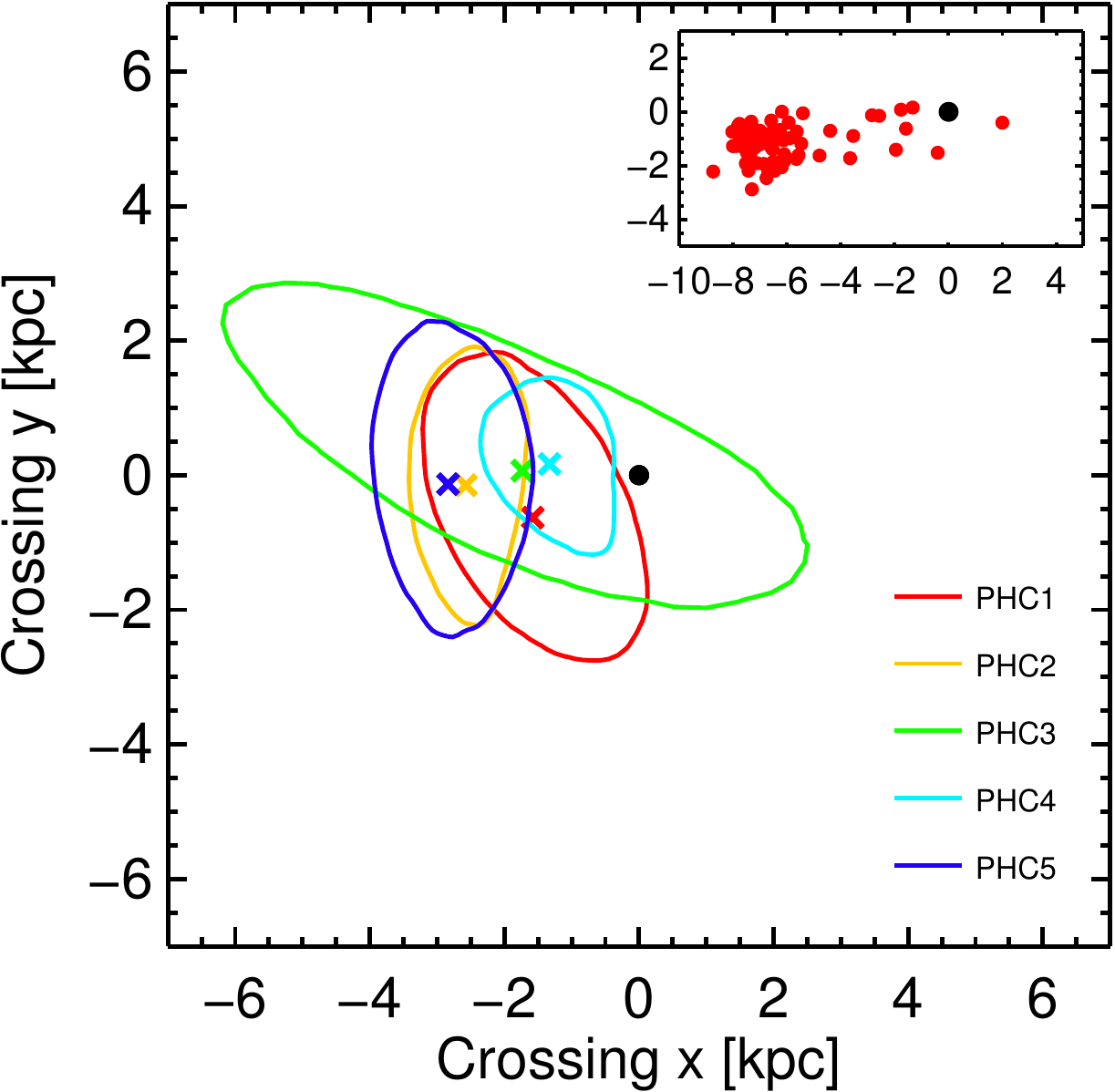}
  \caption{The distribution of last disk crossing location for our
    sample of P200 candidates. The black dot denotes the
    Galactic Center and the red dots in the inset show the crossing
    positions for all targets. The contours show the 
    3$\sigma$ constraints on the crossing location for the five best
    candidates, i.e. with crossing radii less than 3 kpc and $\rm
    [Fe/H]>-0.8$ dex.}
\label{fig:hale_po}
\end{figure}

Given the radial velocity and metallicity from the spectrum, we
first refine our distance estimate using the spectroscopic metallicity
and then compute the crossing position and velocity as before (see
Section \ref{sec:method}). We check the crossing position for the 76
targets and find that 8 have average crossing radii less than 3
kpc, but 3 of these have metallicity less than -0.8. We show the
crossing positions for the remaining 5 metal-rich stars in Figure
\ref{fig:hale_po} and list their properties in Table
\ref{tab:hills_metalrich}. We label these candidates Palomar Hills
Candidates (PHC1--5). The inset of this figure shows the crossing
position of all 76 stars; as expected most stars are on fairly
circular orbits which last crossed the disk close to the solar
radius.

For our best candidate, PHC3, the GC lies within 3$\sigma$ of the
last disk crossing, but the uncertainty on $r_{cr}$ is large.
For PHC4 the 3$\sigma$ likelihood regions is close to the GC, and
both $v_{tot}$ and $v_{cr}$ are large. However, its orbit does not
appear to intersect the GC and so, although this may be a Hills
candidate, it could only be so if there was a systematic error in one
of the observed quantities (e.g. the distance or proper motion).

To compare our proper motion selection algorithm to the blind SDSS
spectroscopic sample, we can see that we have one candidate with
$\rm r_{cr} < 2$ kpc, $\rm P(r_{cr}<1) > 25\%$ and $\rm [Fe/H] >
-0.6$. For our SDSS sample we have 10 stars which pass these
criteria. So the proper motion selected sample has an efficiency of
around $1/76 \sim 1.3\%$, while the SDSS sample has an efficiency of
$10/13170 \sim 0.08\%$. We are in the regime of small
number statistics, but there is undoubtedly a significant improvement
when using our proper motion selected sample.

\section{Possible sources of contamination}
\label{sec:halo}

For all of our Hills candidates, the current total spatial velocity is
around 200 - 300 \kms, which is consistent with typical halo stars. It
is known that halo stars are on radial orbits and so it is natural to
wonder whether they may constitute a significant source of
contamination in our sample. We now address this issue, together with
other possible contaminants at the end of the section.

We can estimate the number of halo stars in our sample of 13,170 SDSS
stars (Cut 4 in Table \ref{tab:number}) by counting the fraction that
are counter-rotating. We find that 1,038 stars are counter-rotating
and, assuming these are all halo stars and that the halo has no significant
net rotation, this implies that the total number of halo stars will be
twice this, i.e. around 16\%. We then estimate the distribution of
peri-centric radii ($\rm r_{peri}$) for a mock sample of halo stars,
using the halo velocity dispersion from \citet{Smith09b}. We take the
positions of these mock halo stars from our observed sample of 13,170
stars, weighted by their halo probability as estimated from the density
distribution of \citet{Juric08}. We Monte Carlo this distribution to
reduce statistical fluctuations and normalize so that the total number
of stars matches the number predicted above.

\begin{figure}
    \plotone{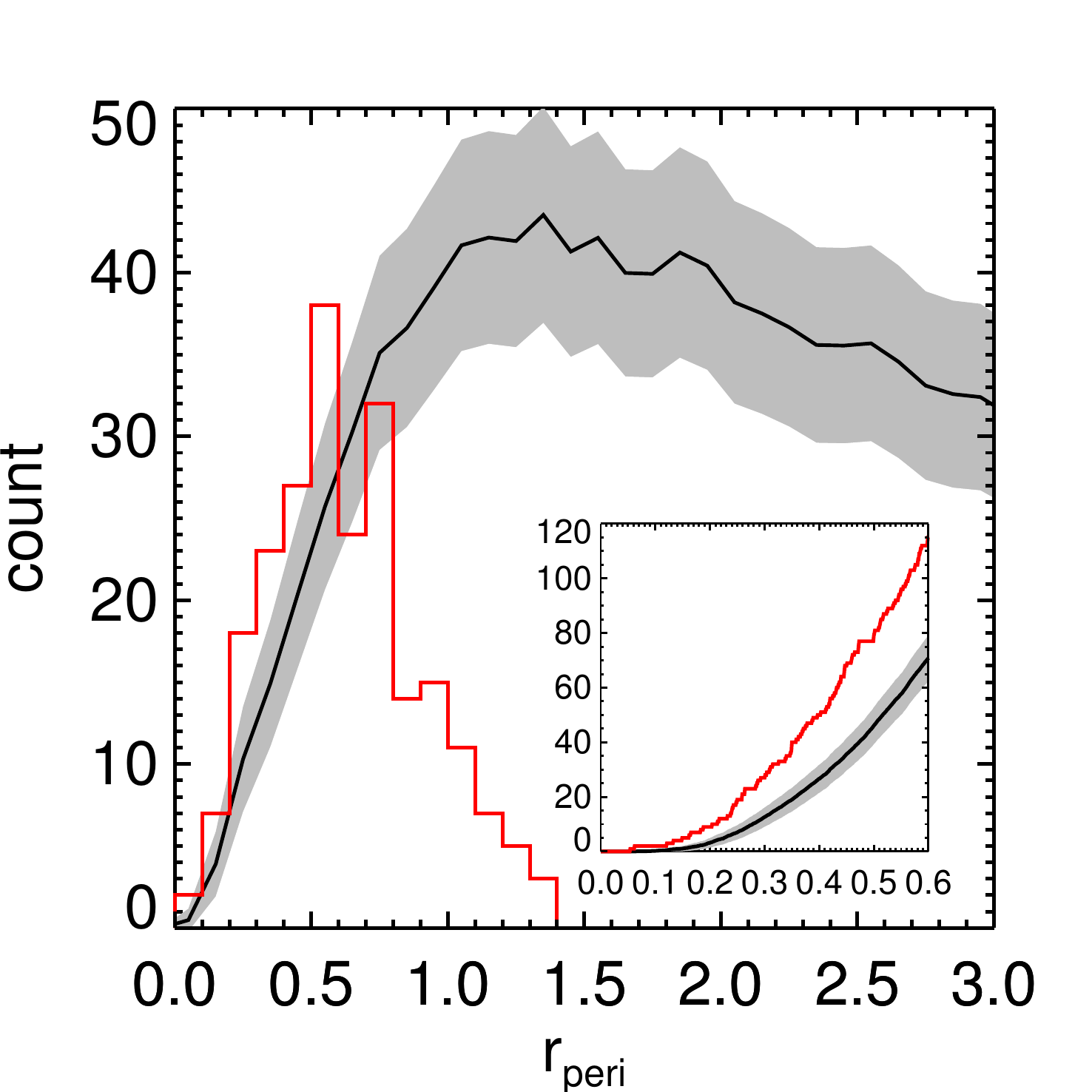}
    \caption{The distribution of the peri-centric radii for our full
    sample of Hills candidates (red line) and our mock halo stars
    (black line). The shaded region shows the 1$\sigma$ uncertainty in
    the mock halo distribution, assuming Poissonian errors. The inset
    shows the cumulative distribution.}
    \label{fig:Hhp}
\end{figure}

The resulting $\rm r_{peri}$ distribution is shown with the black line
in Figure \ref{fig:Hhp}. The shaded region shows the 1$\sigma$
uncertainty in the mock halo distribution, assuming Poissonian errors.
The red line shows the observed distribution for our full sample of
226 Hills candidates, where the fall-off for large $\rm r_{peri}$ is
simply due to the fact that we only considered candidates with small
values of $\rm r_{cr}$. Note that there is a clear excess in the observed
distribution. If there were no Hills stars in this sample, then the
observed number should not exceed the predicted halo distribution, so
this indicates that some of the candidates should be real. This can
also be seen in the inset, which shows the cumulative distribution of
$\rm r_{peri}$; the observed distribution (red curve) is consistently
in excess of the expectations from our mock halo distribution at a
level of around 5$\sigma$.

Although there is a significant excess, there could be deficiencies in
our simplified halo model. Firstly, if the halo exhibits net pro-grade
rotation, then our estimate for the halo fraction will be
underestimated. However, to bridge the gap between the observed and
mock halo distributions would require an unfeasibly high level of
rotation; to increase our halo contribution by around 54\%, which
would bring it into agreement with the observed distribution, the halo
would need to be rotating at around 35 \kms. Although some works have
claimed such high levels of rotation may be possible
\citep[e.g.][]{Deason2011}, the general belief is that rotation should
not be this large
\citep[e.g.][]{Allende2006,Smith09a,Bond2010,Fermani2013,King2015}. The second
potential deficiency in the halo model is the level of
anisotropy. However, our model adopts the dispersions of
\citet{Smith09a} with spherical anisotropy parameter,
\begin{equation}
  \beta = 1 - \frac{\sigma^2_\phi + \sigma^2_\theta}{2\sigma^2_r}= 0.69,
\end{equation}
which is at the radial end of the commonly assumed values
\citep[measurements of $\beta$ typically range from around 0.4 to
0.6;][]{Smith09a,Williams2015,King2015,Das2016}. Therefore, although
we concede that the halo model parameters could be tweaked in order to
decrease the discrepancy between the predicted distributions of $\rm
r_{peri}$, such an approach is unlikely to completely reconcile the
discrepancy.

If we accept that halo contamination may be present, one way to
discriminate against this is to impose a metallicity cut, as we have
discussed above. By restricting our sample to $\rm [Fe/H] > -0.6$ we
remove a large amount of the halo contamination (remember that the
halo has a mean $-1.46$ dex and dispersion $0.3$ dex;
\citealt{Ivezic08}) and should retain any Hills stars, which are
expected to have high metallicities.

\begin{figure}
    \plotone{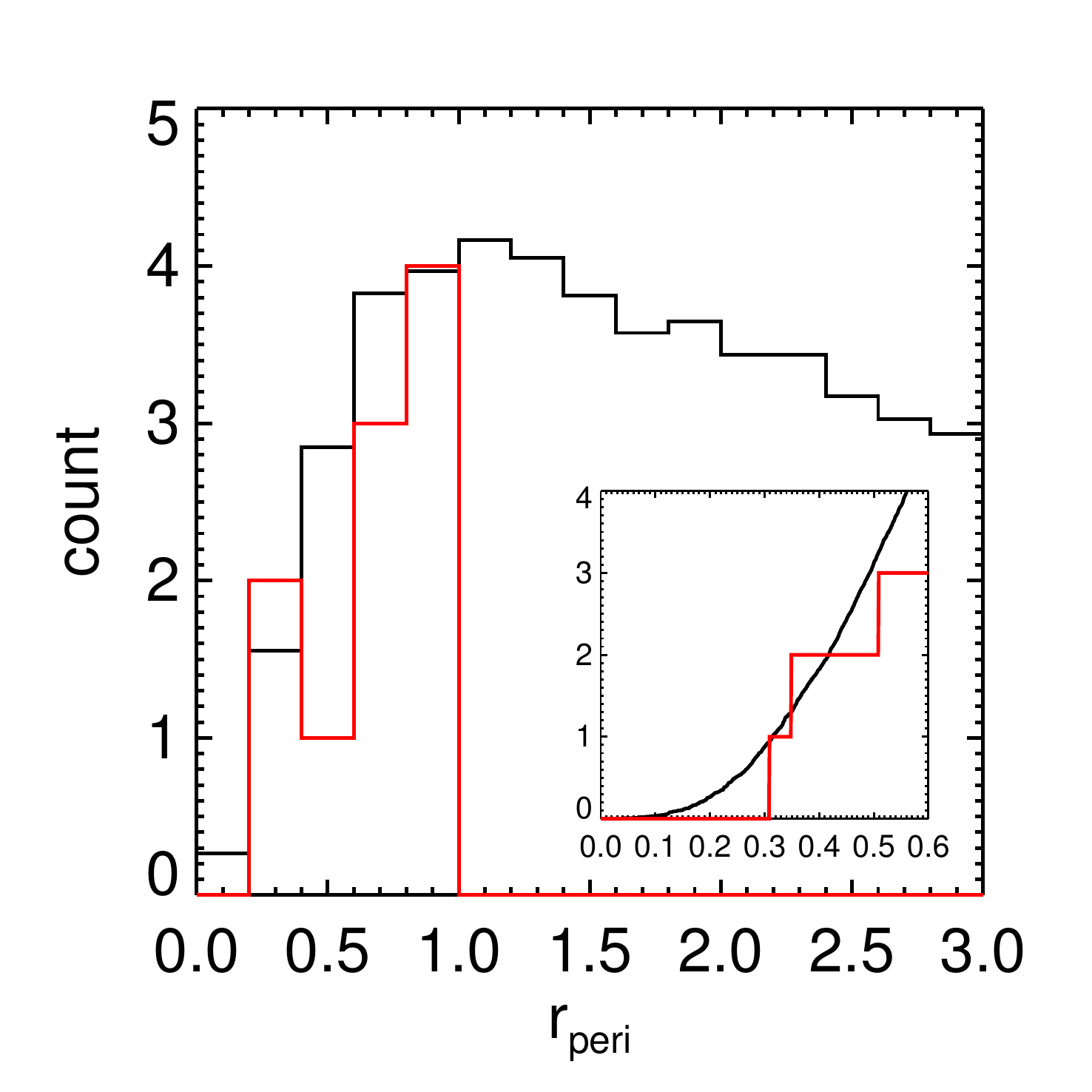}
    \caption{As Figure \ref{fig:Hhp}, but for our metal-rich sample of
      Hills candidates ($\rm [Fe/H] > -0.6$).}
    \label{fig:Hhpmr}
\end{figure}

This reduces our sample considerably. The total number of spectra
falls from 13,170 to 4,893 and the number of Hills candidates is only
10. Despite these small numbers, we can repeat the above exercise and
predict the distribution of $\rm r_{peri}$ for the halo. There are
only 48 counter-rotating stars in this metallicity range,
indicating that the halo fraction is around 1.0\%. The resulting
prediction for the halo $\rm r_{peri}$ distribution is shown by the
black line in Figure \ref{fig:Hhpmr} and can be compared to the
observed distribution in red. Here the agreement is much better,
indicating that these stars are more consistent with being from the
metal-rich tail of the halo population.

It is interesting to note the discrepancy between the two
distributions in Figure \ref{fig:Hhp} is not reflected in the
metal-rich sample in Figure \ref{fig:Hhpmr}. For example, the full
sample of Hills candidates has 77 stars with $\rm r_{peri}<0.5$ kpc,
while our mock halo sample predicts only 50, but for the metal-rich
sample only 2 stars remain with $\rm r_{peri}<0.5$ kpc. This leaves
25 metal-poor stars unaccounted for and, as mentioned above,
we  do not believe this can be explained by deficiencies in our halo
model.

Two other possible sources of contamination are bulge and
thick-disk stars on radial orbits, but neither of these are expected
to be significant.
As pointed out by \citet{Vickers15}, the bulge dispersion of around
100 \kms is insufficient to produce a significant number of stars in
the solar neighborhood (note that our sample are predicted to have
left the bulge with velocities in excess of 450 \kms).
We estimate the thick disk contamination using the Besancon model
\citet{Robin2003}. This kinematics in this model are flawed in that it
assumes a Gaussian profile for the azimuthal velocities, whereas in
reality the asymmetric drift induces a skew in the distribution
\citep[see Section 4.8.2 of][]{BT08}. To alleviate this issue, we
replace Besancon's azimuthal velocity distribution with the
physically-motivated distribution of \citet{Schonrich2012b}. This 
model predicts negligible thick-disk contamination. Although it is not
inconceivable that one thick-disk star may be present in our two
samples, the chances of significant contamination is small. For
example, for the full sample of 13,170 stars, the probability that we
have one thick-disk star on a highly radial orbit ($\rm r_{peri}<0.5$
kpc) is 27\%, but the probability that we have three or more is less
than 1\%. For the metal-rich sample of 4,893 stars, the probability of
obtaining one radial thick-disk star is 8\%, but less than 1\% for
two or more stars.

In conclusion, although our metal-rich Hills candidates appear to be
consistent with predictions for halo contamination (albeit in some
cases with unusually high metallicities), the metal-poor candidates
are in conflict.
While the nature of our Hills candidates is unclear, it is unlikely
that the metal-poor candidates can be explained away by
halo, bulge or disk contamination. One could argue that they are the
tail of high velocity runaway stars, ejected from mechanisms other
than an interaction with the supermassive black hole. However,
assuming these stars originated in the bulge, our ejection velocities
are all in excess of 400 \kms, which is hard to achieve by standard
runaway mechanisms (see Section \ref{sec:intro}).
The more metal-rich stars could potentially have been ejected
from the disk, but this would require an ejection velocity vector
orientated precisely against the disk's rotation in order to place the
star on a highly radial orbit. This is possible, but the combination
of (a) the requirement on the ejection vector, together with the fact
that (b) such ejection speeds are towards the tail of the predicted
distribution, mean that we do not expect this to be a major source of
contamination.

\section{Conclusion}
\label{sec:conclusion}

Stars ejected from the Galactic centre can be used to place important
constraints on the Milky Way potential \citep[e.g.][]{Gnedin2005}.
Although many such stars have been found, most are too distant to
obtain the detailed 6D phase-space information that is required to
constrain the potential. We have therefore carried out a search for
nearby ejected stars, focusing not only on unbound ejecta
(hypervelocity stars) but also including bound ejecta (which we dub
Hills stars). By integrating orbits back through time we have
identified a number of candidates, with orbits consistent with
Galactic centre ejection. Our analysis used data from SDSS and, since
the efficacy of such studies are often hampered by deficiencies in
proper motion catalogs, we have chosen to utilize the reliable,
high-precision Stripe 82 proper motion catalogue
\citep{Bramich08,Koposov13}.

From the 13,170 SDSS stellar spectra in Stripe 82, we found 226
candidate Hills stars. As was discussed in Section \ref{sec:halo},
this number is significantly in excess of what we would expect for
halo stars on radial orbits and cannot be explained by disk or bulge
contamination. Deficiencies in our halo model, due to incorrect
assumptions about rotation or anisotropy, are unlikely to reconcile
this discrepancy. One might speculate that an accretion event on a
very radial orbit could cause this signal, but in conclusion the
nature of the metal-poor candidates remains uncertain.

Since all known hypervelocity stars to-date are metal rich, we
expect that our Hills candidates should also be metal rich. However,
there is a selection bias at play, since hypervelocity stars have
been identified based on the fact that they are young (and hence
metal-rich) stars that stand out amongst the old populations of the
Galactic halo. So it is possible that there may be low-mass metal-poor
hypervelocity stars, but they have remained hidden from the existing
search techniques. On the other hand, because the Galactic bulge was enriched
on such a short time-scale, the fraction of metal-poor stars in the
bulge will be small, even at earlier times. This indicates that it
would be surprising if we were to find a significant population of
metal-poor Hills stars.
If we follow the assumption that Hills stars should be metal-rich and
restrict ourselves to such stars, we find 29 candidates with
$\rm [Fe/H] > -0.8$ dex and 10 with $\rm [Fe/H] > -0.6$ dex. Their
metallicities are more consistent with what we expect for bulge
ejecta, and so follow-up observations could be particularly
interesting for these stars. It is worth noting that even if these are
genuine Hills stars, none of them are unbound and hence our result is
in agreement with the non-detection from \citet{Kollmeier10}.

We can carry out an order-of-magnitude estimate of the
ejection rate for these candidates, following
\citet[][see the first paragraph of Section 3.2]{Kollmeier10}. The
survey volume of our 250 sq. deg. wedge is around 3.2 kpc$^3$, assuming we
probe distances of 0.5 to 5 kpc. However, from Table \ref{tab:number}
we can see only 5.7\% of stars have spectra, meaning that the
effective volume is only 0.18 kpc$^3$. This is equivalent to a thin
Galacto-centric shell at 9 kpc of width $1.7\times10^{-4}$ kpc, which
a 300 km/s Hills star will spend $5.7\times10^{-4}$ Myr
traversing. Therefore, if we assume we have between 1 to 10 Hills stars
in our SDSS sample, this implies an ejection rate of
$2-20\times10^3$ Myr$^{-1}$. This is a very simplified calculation which
neglects two important considerations: firstly, the survey volume we
probe depends on the spectral type; and secondly, since our sample are
all bound, they may not be on their first passage through the solar
radius. Despite these limitations, our estimate is comparable to
existing predictions; although classical B-type HVSs are predicted to
be ejected at a rate of 100 to 1000 Myr$^{-1}$
\citep[e.g.][]{Zhang13}, \citet{Kenyon08} calculate that the ejection
rate of low-mass Hills stars could be around 2 to 3 orders-of-magnitude
greater than that of B-type stars. Clearly larger samples with better
statistics are required to fully address the question of ejection rates.

To address the issue of sample size, we have developed a
technique using proper-motions to optimize candidate selection
(Section \ref{sec:palomar}). We have tested this approach with
our own radial velocities obtained from the Palomar 200-inch
telescope. This technique provides considerable improvement on the
blind spectroscopic sample of SDSS, being able to identify Hills
candidates with an efficiency around 20 times better than a blind
search. We believe there is good potential for this approach in
future. At 250 sq. deg., Stripe 82 covers less than 150th of the total
sky. Current surveys, such as Pan-STARRS \citep{Kaiser2010}, could
immediately be used to repeat this exercise for significantly larger
volumes.

\begin{figure}
  \plotone{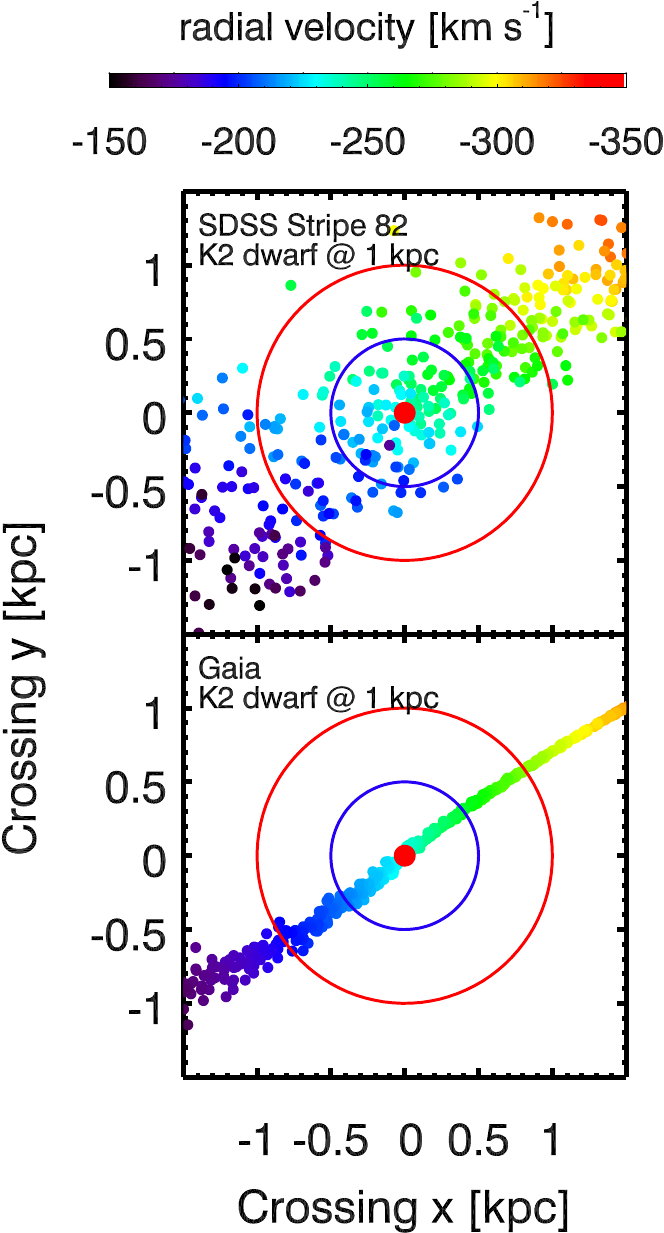}
  \caption{
An illustration of how our proper motion selection technique will
improve with forthcoming Gaia data. We show two disk crossing tracks for a
hypothetical Hills star, assuming this is a K2-dwarf located at 1
kpc. The top panel shows typical Stripe 82 uncertainties (errors of
20\% in distance and 1.5 $\rm mas/yr$ in proper motion), while the
bottom panel shows projected end-of-mission Gaia errors (4.5\% in
distance and 0.02 $\rm mas/yr$ in proper motion).}
\label{fig:vrad_gaia}
\end{figure}

Clearly the efficacy of this method depends on the uncertainties 
in the observed quantities, most notably the distances and proper
motions. The thickness of the disk crossing track in Figure \ref{fig:vrad}
(see Section \ref{sec:target_selection} determines how reliable our
method is; if we knew all observed photometric properties to infinite
precision, then this track would be a line and for bona-fide Hills
stars this would intersect the GC. The Gaia mission will provide both
distances and proper motions to exquisite precision for many millions
of stars in the solar neighborhood, together with metallicities from
the on-board spectro-photometry\footnote{Although radial velocities
will also be obtained, this is only for the brightest stars, obtaining
accuracies of around 5 \kms at $V \sim 14$ mag. See
\url{http://www.cosmos.esa.int/web/gaia/science-performance} and
\citet{deBruijne12}} \citep{Liu12}.
Such a dataset would dramatically improve the efficacy of our method
(see Fig. \ref{fig:vrad_gaia}), producing a catalogue of high-probability
candidates that could be followed up with moderate observing resources. 
The accurate distances and proper motions leads to much more precise
knowledge of the ejected star's orbit, consequently providing greatly
improved constraints on the Galactic potential.
We are now working on a follow-up paper to quantify how many Hills
stars would be required to place meaningful constraints on the
potential and this will be reported in due course.

\section*{Acknowledgments}

The authors wish to thank Alberto Rebassa-Mansergas, Matthew Molloy,
Eric Peng, and Hengxiao Guo for assistance with this work, and
Jinliang Hou and the cluster group at SHAO for useful discussions.
This work is supported by the CAS One Hundred Talent Fund, NSFC
Grants 11173002 \& 11333003, the Strategic Priority Research Program
The Emergence of Cosmological Structures of the Chinese Academy of
Sciences (XDB09000000), the National Key Basic Research Program of
China (2014CB845700) and the National Science Foundation grant AST
14-09421.

It uses data obtained through the Telescope Access Program (TAP),
which has been funded by the Strategic Priority Research Program ”The
Emergence of Cosmological Structures” (Grant No. XDB09000000),
National Astronomical Observatories, Chinese Academy of Sciences, and
the Special Fund for Astronomy from the Ministry of
Finance. Observations obtained with the Hale Telescope at Palomar
Observatory were obtained as part of an agreement between the National
Astronomical Observatories, Chinese Academy of Sciences, and the
California Institute of Technology. Observations obtained with the
Hale Telescope at Palomar Observatory were obtained as part of an
agreement between the National Astronomical Observatories, Chinese
Academy of Sciences, and the California Institute of Technology.

\appendix

\section{Additional Hills candidates from SDSS spectroscopy}

\begin{figure*}
\begin{center}
\includegraphics[width=18cm]{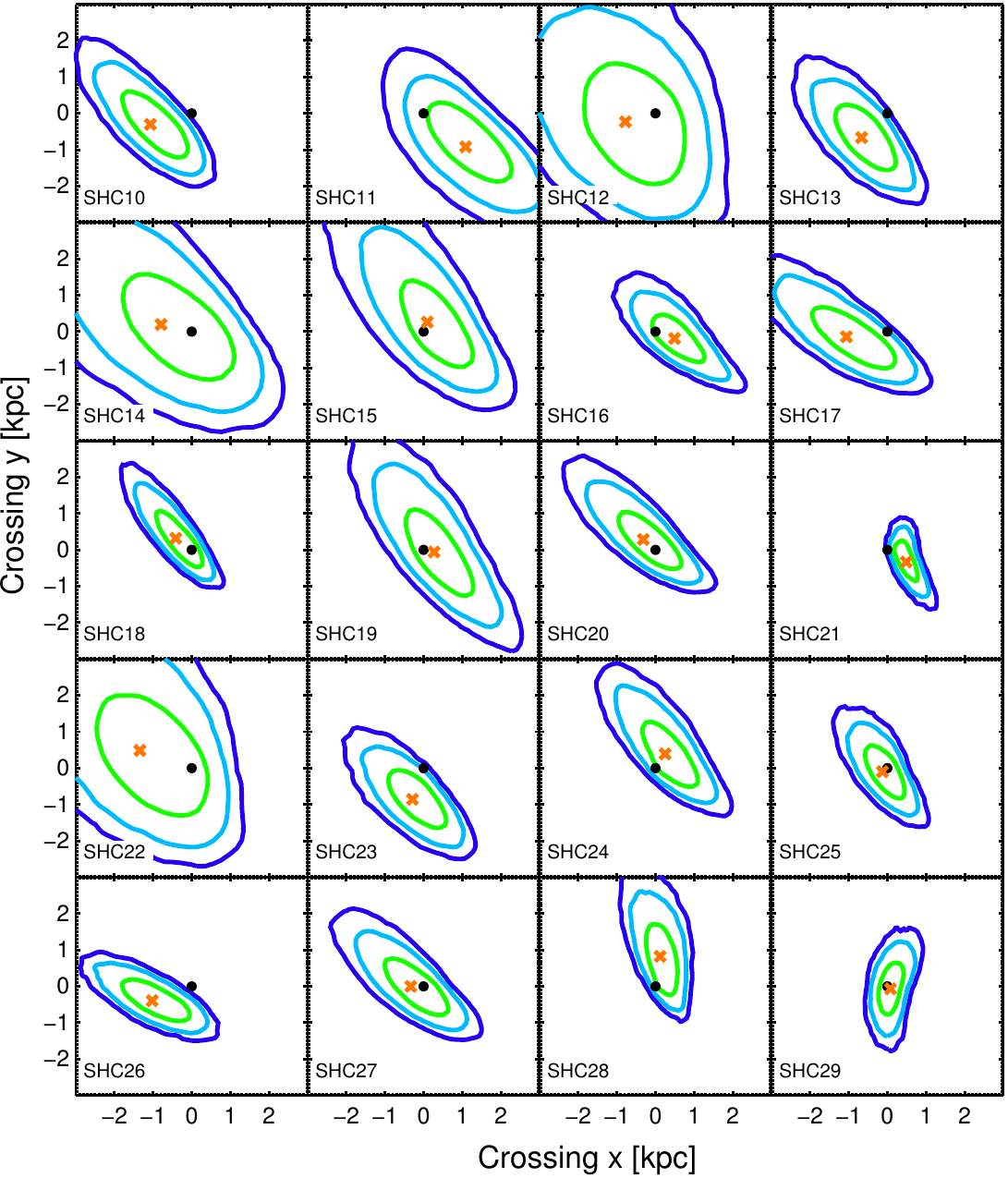}
  \caption{This figure shows the last disk crossing position for our
    moderately metal-rich SDSS Hills 
    candidates (i.e. with $-0.8 \leqslant [Fe/H] \leqslant -0.59$ dex).
    The green, cyan and blue lines correspond to
    1$\sigma$, 2$\sigma$ and 3$\sigma$, respectively. The orange
    crosses show the crossing position if we do not account for
    observational errors.
    The black dot shows the position of Galactic center. The nine most
    metal-rich candidates are presented in Fig. \ref{fig:hills_metalrich}.}
    \label{fig:hills_candidates}
\end{center}
\end{figure*}

Here, in Figure \ref{fig:hills_candidates}, we present plots of the
crossing position for the remaining Hills candidates from Table
\ref{tab:hills_metalrich}. The best nine candidates (i.e. with $\rm
[Fe/H]\geqslant-0.57$ dex) are presented in Fig. \ref{fig:hills_metalrich}.

\end{document}